\documentclass[review]{elsarticle}

\usepackage{lineno,hyperref}
\modulolinenumbers[5]

\journal{Neural Networks}
\usepackage[english]{babel}
\usepackage{graphicx}
\usepackage{subfigure}
\usepackage{amssymb}
\usepackage{amsmath}

\makeatletter
\def\hlinewd#1{%
  \noalign{\ifnum0=`}\fi\hrule \@height #1 \futurelet
   \reserved@a\@xhline}
\makeatother
\usepackage{multirow}

\begin{document}

\begin{frontmatter}

\title{Self-Taught Convolutional Neural Networks for\\Short Text Clustering}


\author[mymainaddress]{Jiaming Xu}
\author[mymainaddress]{Bo Xu\corref{mycorrespondingauthor}}
\author[mymainaddress]{Peng Wang}
\author[mymainaddress]{Suncong Zheng}
\cortext[mycorrespondingauthor]{Corresponding author. Email address: boxu@ia.ac.cn}
\author[mymainaddress]{\\Guanhua Tian}
\author[mymainaddress,mysecondaryaddress]{Jun Zhao}

\author[mymainaddress,brainaddress]{Bo Xu}

\address[mymainaddress]{Institute of Automation, Chinese Academy of Sciences (CAS), Beijing, P.R. China}
\address[mysecondaryaddress]{National Laboratory of Pattern Recognition (NLPR), Beijing, P.R. China}
\address[brainaddress]{Center for Excellence in Brain Science and Intelligence Technology, CAS. P.R. China}

\begin{abstract}
  Short text clustering is a challenging problem due to its sparseness of text representation. Here we propose a flexible Self-Taught Convolutional neural network framework for Short Text Clustering (dubbed STC$^2$), which can flexibly and successfully incorporate more useful semantic features and learn non-biased deep text representation in an unsupervised manner. In our framework, the original raw text features are firstly embedded into compact binary codes by using one existing unsupervised dimensionality reduction methods. Then, word embeddings are explored and fed into convolutional neural networks to learn deep feature representations, meanwhile the output units are used to fit the pre-trained binary codes in the training process. Finally, we get the optimal clusters by employing K-means to cluster the learned representations. Extensive experimental results demonstrate that the proposed framework is effective, flexible and outperform several popular clustering methods when tested on three public short text datasets.
\end{abstract}


\end{frontmatter}


\section{Introduction}
\label{sec:Introduction}
Short text clustering is of great importance due to its various applications, such as user profiling~\cite{li-ritter-hovy:2014:P14-1} and recommendation~\cite{wang-EtAl:2010:ACL1}, for nowaday's social media dataset emerged day by day. However, short text clustering has the data sparsity problem and most words only occur once in each short text~\cite{15_aggarwal2012survey}. As a result, the Term Frequency-Inverse Document Frequency (TF-IDF) measure cannot work well in short text setting. In order to address this problem, some researchers work on expanding and enriching the context of data from Wikipedia~\cite{29_banerjee2007clustering} or an ontology~\cite{33_fodeh2011ontology}. However, these methods involve solid Natural Language Processing (NLP) knowledge and still use high-dimensional representation which may result in a waste of both memory and computation time. Another way to overcome these issues is to explore some sophisticated models to cluster short texts. For example, Yin and Wang~\cite{30_yin2014dirichlet} proposed a Dirichlet multinomial mixture model-based approach for short text clustering. Yet how to design an effective model is an open question, and most of these methods directly trained based on Bag-of-Words (BoW) are shallow structures which cannot preserve the accurate semantic similarities.

Recently, with the help of word embedding, neural networks demonstrate their great performance in terms of constructing text representation, such as Recursive Neural Network (RecNN)~\cite{24_socher2011semi,35_socher2013recursive} and Recurrent Neural Network (RNN)~\cite{38_mikolov2011extensions}. However, RecNN exhibits high time complexity to construct the textual tree, and RNN, using the hidden layer computed at the last word to represent the text, is a biased model where later words are more dominant than earlier words~\cite{14_lai2015rcnn}. Whereas for the non-biased models, the learned representation of one text can be extracted from all the words in the text with non-dominant learned weights. More recently, Convolution Neural Network (CNN), as the most popular non-biased model and applying convolutional filters to capture local features, has achieved a better performance in many NLP applications, such as sentence modeling~\cite{16_blunsom2014convolutional}, relation classification~\cite{34_zeng2014relation}, and other traditional NLP tasks~\cite{19_collobert2011natural}. Most of the previous works focus CNN on solving supervised NLP tasks, while in this paper we aim to explore the power of CNN on one unsupervised NLP task, short text clustering.

\begin{figure}[t]
\begin{center}
\includegraphics[width=7.7cm]{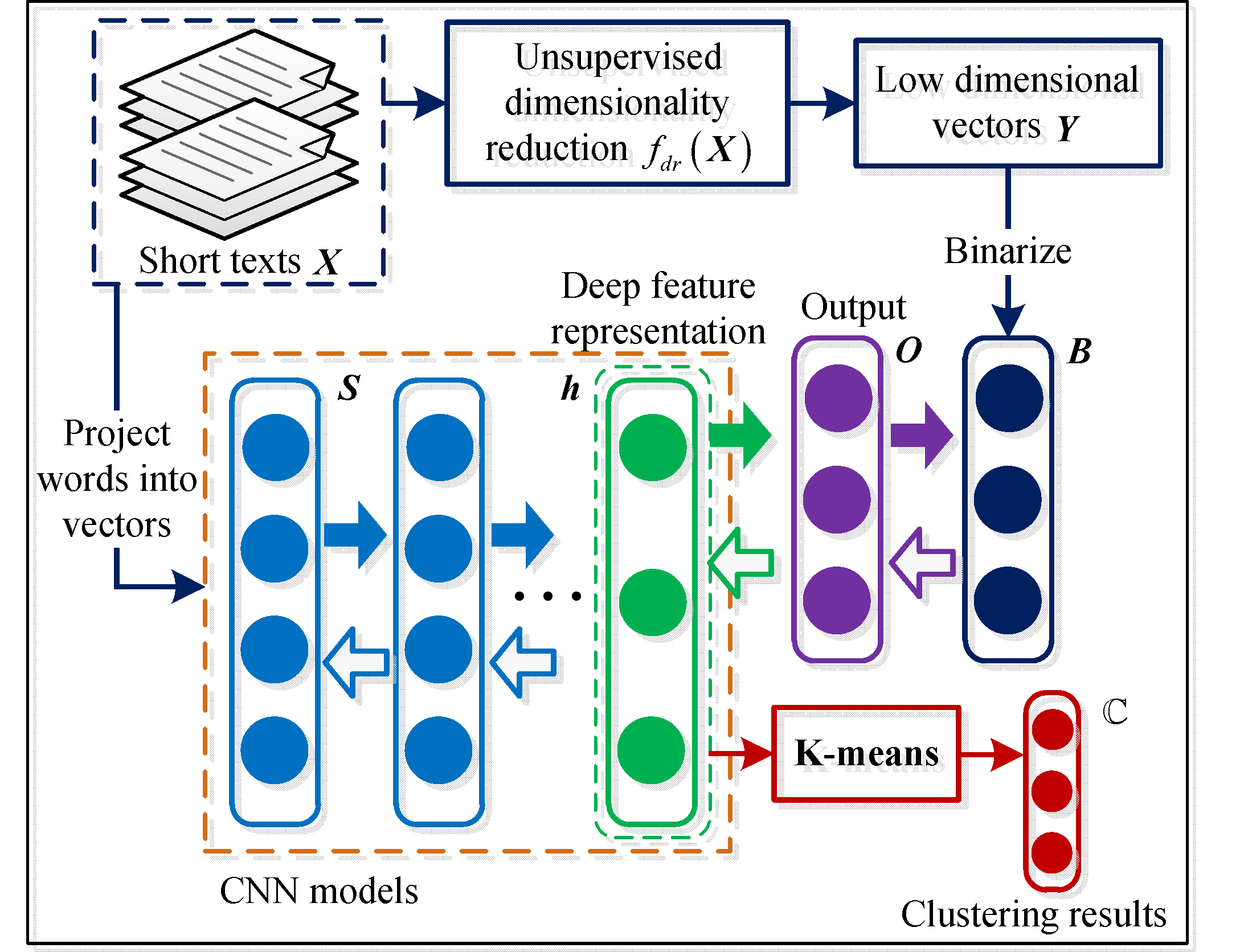}
\caption{The architecture of our proposed STC$^2$ framework for short text clustering. Solid and hollow arrows represent forward and backward propagation directions of features and gradients respectively. The STC$^2$ framework consist of deep convolutional neural network (CNN), unsupervised dimensionality reduction function and K-means module on the deep feature representation from the top hidden layers of CNN.}\label{fig:Architecture}
\end{center}
\end{figure}

We systematically introduce a simple yet surprisingly powerful Self-Taught Convolutional neural network framework for Short Text Clustering, called STC$^2$. An overall architecture of our proposed approach is illustrated in Figure~\ref{fig:Architecture}. We, inspired by~\cite{28_zhang2010self,TwoStepHash_Lin_2013}, utilize a self-taught learning framework into our task. In particular, the original raw text features are first embedded into compact binary codes \({\bf{B}}\) with the help of one traditional unsupervised dimensionality reduction function. Then text matrix \({\bf{S}}\) projected from word embeddings are fed into CNN model to learn the deep feature representation \({\bf{h}}\) and the output units are used to fit the pre-trained binary codes \({\bf{B}}\). After obtaining the learned features, K-means algorithm is employed on them to cluster texts into clusters \(\mathbb{C}\). Obviously, we call our approach ``self-taught'' because the CNN model is learnt from the pseudo labels generated from the previous stage, which is quite different from the term ``self-taught'' in~\cite{raina2007self}.
Our main contributions can be summarized as follows:

\begin{itemize}
  \item We propose a flexible short text clustering framework which explores the feasibility and effectiveness of combining CNN and traditional unsupervised dimensionality reduction methods.
  \item Non-biased deep feature representations can be learned through our self-taught CNN framework which does not use any external tags/labels or complicated NLP pre-processing.
  \item We conduct extensive experiments on three short text datasets. The experimental results demonstrate that the proposed method achieves excellent performance in terms of both accuracy and normalized mutual information.
\end{itemize}

This work is an extension of our conference paper~\cite{xu2015short}, and they differ in the following aspects. First, we put forward a general a self-taught CNN framework in this paper which can flexibly couple various semantic features, whereas the conference version can be seen as a specific example of this work. Second, in this paper we use a new short text dataset, Biomedical, in the experiment to verify the effectiveness of our approach. Third, we put much effort on studying the influence of various different semantic features integrated in our self-taught CNN framework, which is not involved in the conference paper.

For the purpose of reproducibility, we make the datasets and software used in our experiments publicly available at the website\footnote{Our code and dataset are available: \url{https://github.com/jacoxu/STC2}}.

The remainder of this paper is organized as follows: In Section~\ref{sec:RelatedWork}, we first briefly survey several related works. In Section~\ref{sec:Methodology}, we describe the proposed approach STC$^2$ and implementation details. Experimental results and analyses are presented in Section~\ref{sec:Experiments}. Finally, conclusions are given in the last Section.

\section{Related Work}
\label{sec:RelatedWork}
In this section, we review the related work from the following two perspectives: short text clustering and deep neural networks.
\subsection{Short Text Clustering}
There have been several studies that attempted to overcome the sparseness of short text representation. One way is to expand and enrich the context of data. For example, Banerjee et al.~\cite{29_banerjee2007clustering} proposed a method of improving the accuracy of short text clustering by enriching their representation with additional features from Wikipedia, and Fodeh et al.~\cite{33_fodeh2011ontology} incorporate semantic knowledge from an ontology into text clustering. However, these works need solid NLP knowledge and still use high-dimensional representation which may result in a waste of both memory and
computation time. Another direction is to map the original features into reduced space, such as Latent Semantic Analysis (LSA)~\cite{deerwester1990indexing}, Laplacian Eigenmaps (LE)~\cite{37_ng2002spectral}, and Locality Preserving Indexing (LPI)~\cite{niyogi2004locality}. Even some researchers explored some sophisticated models to cluster short texts. For example, Yin and Wang~\cite{30_yin2014dirichlet} proposed a Dirichlet multinomial mixture model-based approach for short text clustering. Moreover, some studies even focus the above both two streams. For example, Tang et al.~\cite{32_tang2012enriching} proposed a novel framework which enrich the text features by employing machine translation and reduce the original features simultaneously through matrix factorization techniques.

Despite the above clustering methods can alleviate sparseness of short text representation to some extent, most of them ignore word order in the text and belong to shallow structures which can not fully capture accurate semantic similarities.

\subsection{Deep Neural Networks}
Recently, there is a revival of interest in DNN and many researchers have concentrated on using Deep Learning to learn features. Hinton and Salakhutdinov~\cite{23_hinton2006reducing} use DAE to learn text representation. During the fine-tuning procedure, they use backpropagation to find codes that are good at reconstructing the word-count vector.

More recently, researchers propose to use external corpus to learn a distributed representation for each word, called word embedding~\cite{25_turian2010word}, to improve DNN performance on NLP tasks. The Skip-gram and continuous bag-of-words models of Word2vec~\cite{21_mikolov2013distributed} propose a simple single-layer architecture based on the inner product between two word vectors, and Pennington et al.~\cite{26_pennington2014glove} introduce a new model for word representation, called GloVe, which captures the global corpus statistics.

In order to learn the compact representation vectors of sentences, Le and Mikolov~\cite{le2014distributed} directly extend the previous Word2vec~\cite{21_mikolov2013distributed} by predicting words in the sentence, which is named Paragraph Vector (Para2vec). Para2vec is still a shallow window-based method and need a larger corpus to yield better performance. More neural networks utilize word embedding to capture true meaningful syntactic and semantic regularities, such as RecNN~\cite{24_socher2011semi,35_socher2013recursive} and RNN~\cite{38_mikolov2011extensions}. However, RecNN exhibits high time complexity to construct the textual tree, and RNN, using the layer computed at the last word to represent the text, is a biased model. Recently, Long Short-Term Memory (LSTM)~\cite{hochreiter1997long} and Gated Recurrent Unit (GRU)~\cite{cho2014learning}, as sophisticated recurrent hidden units of RNN, has presented its advantages in many sequence generation problem, such as machine translation~\cite{sutskever2014sequence}, speech recognition~\cite{graves2013speech}, and text conversation~\cite{shang2015neural}. While, CNN is better to learn non-biased implicit features which has been successfully exploited for many supervised NLP learning tasks as described in Section~\ref{sec:Introduction}, and various CNN based variants are proposed in the recent works, such as Dynamic Convolutional Neural Network (DCNN)~\cite{16_blunsom2014convolutional}, Gated Recursive Convolutional Neural Network (grConv)~\cite{cho2014properties} and Self-Adaptive Hierarchical Sentence model (AdaSent)~\cite{zhao2015self}.

In the past few days, Visin et al.~\cite{visin2015renet} have attempted to replace convolutional layer in CNN to learn non-biased features for object recognition with four RNNs, called ReNet, that sweep over lower-layer features in different directions: (1) bottom to top, (2) top to bottom, (3) left to right and (4) right to left. However, ReNet does not outperform state-of-the-art convolutional neural networks on any of the three benchmark datasets, and it is also a supervised learning model for classification. Inspired by Skip-gram of word2vec~\cite{mikolov2013efficient,21_mikolov2013distributed}, Skip-thought model~\cite{kiros2015skip} describe an approach for unsupervised learning of a generic, distributed sentence encoder. Similar as Skip-gram model, Skip-thought model trains an encoder-decoder model that tries to reconstruct the surrounding sentences of an encoded sentence and released an off-the-shelf encoder to extract sentence representation. Even some researchers introduce continuous Skip-gram and negative sampling to CNN for learning visual representation in an unsupervised manner~\cite{wang2015unsupervised}.
This paper, from a new perspective, puts forward a general self-taught CNN framework which can flexibly couple various semantic features and achieve a good performance on one unsupervised learning task, short text clustering.


\section{Methodology}
\label{sec:Methodology}
  Assume that we are given a dataset of \(n\) training texts denoted as: \({\bf{X}} = {\{ {{\bf{x}}_i}:{{\bf{x}}_i} \in {\mathbb{R}^{d \times 1}}\} _{i = 1,2,...,n}}\), where \(d\) is the dimensionality of the original BoW representation. Denote its tag set as \({\bf{T}} = \{ 1,2,...C\} \) and the pre-trained word embedding set as \({\bf{E}} = {\{ {\bf{e}}({w_i}):{\bf{e}}({w_i}) \in {\mathbb{R}^{{d_w} \times 1}}\} _{i = 1,2,...,|V|}}\), where \(d_w\) is the dimensionality of word vectors and $|V|$ is the vocabulary size. In order to learn the \(r\)-dimensional deep feature representation \({\bf{h}}\) from CNN in an unsupervised manner, some unsupervised dimensionality reduction methods \({f_{dr}}\left( {\bf{X}} \right)\) are employed to guide the learning of CNN model. Our goal is to cluster these texts \({\bf{X}}\) into clusters \(\mathbb{C}\) based on the learned deep feature representation while preserving the semantic consistency.

As depicted in Figure~\ref{fig:Architecture}, the proposed framework consist of three components, deep convolutional neural network (CNN), unsupervised dimensionality reduction function and K-means module. In the rest sections, we first present the first two components respectively, and then give the trainable parameters and the objective function to learn the deep feature representation. Finally, the last section describe how to perform clustering on the learned features.

\subsection{Deep Convolutional Neural Networks}

\begin{figure}[t]
\begin{center}
\includegraphics[width=7.7cm]{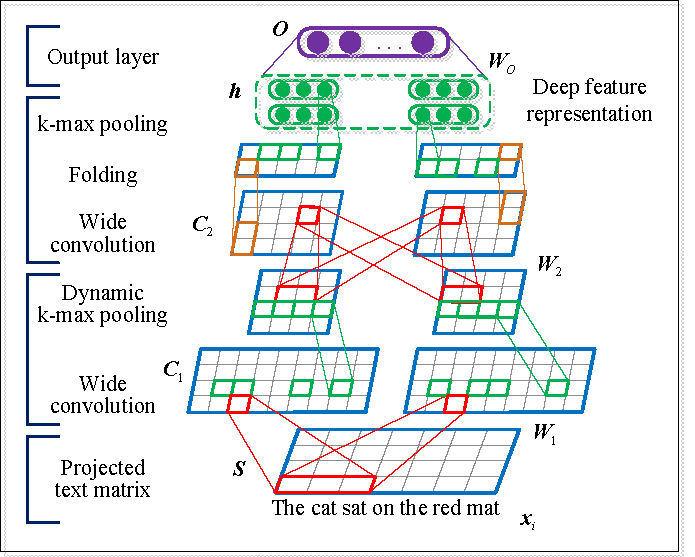}
\caption{The architecture of dynamic convolutional neural network~\cite{16_blunsom2014convolutional}. An input text is first projected to a matrix feature by looking up word embeddings, and then goes through wide convolutional layers, folding layers and k-max pooling layers, which provides a deep feature representation before the output layer.}\label{fig:DCNN}
\end{center}
\end{figure}


In this section, we briefly review one popular deep convolutional neural network, Dynamic Convolutional Neural Network (DCNN)~\cite{16_blunsom2014convolutional} as an instance of CNN in the following sections, which as the foundation of our proposed method has been successfully proposed for the completely supervised learning task, text classification.

Taking a neural network with two convolutional layers in Figure~\ref{fig:DCNN} as an example, the network transforms raw input text to a powerful representation. Particularly, each raw text vector \({{\bf{x}}_i}\) is projected into a matrix representation \({\bf{S}} \in {\mathbb{R}^{{d_w} \times s}}\) by looking up a word embedding \({\bf{E}}\), where \({s}\) is the length of one text. We also let \({\bf{\tilde W}} = {\{ {{\bf{W}}_i}\} _{i = 1,2}}\) and \({{\bf{W}}_O}\) denote the weights of the neural networks. The network defines a transformation \(f( \cdot ):{\mathbb{R}^{d \times 1}} \to {\mathbb{R}^{r \times 1}}\) \((d \gg r)\) which transforms an input raw text \({\bf{x}}\) to a \(r\)-dimensional deep representation \({\bf{h}}\). There are three basic operations described as follows:

 {\bf{Wide one-dimensional convolution}} This operation \({\bf{m}} \in {\mathbb{R}^m}\) is applied to an individual row of the sentence matrix \({\bf{S}} \in {\mathbb{R}^{{d_w} \times s}}\), and yields a resulting matrix \({\bf{C}} \in {\mathbb{R}^{  {d_w} \times {(s + m - 1)} }}\), where \(m\) is the width of convolutional filter.

 {\bf{Folding}}  In this operation, every two rows in a feature map are simply summed component-wisely. For a map of \(d_w\) rows, folding returns a map of \(d_w/2\) rows, thus halving the size of the representation and yielding a matrix feature \({\bf{\hat C}} \in {\mathbb{R}^{  {(d_w/2)} \times {(s + m - 1)} }}\). Note that folding operation does not introduce any additional parameters.

{\bf{Dynamic \(k\)-max pooling}} Assuming the pooling parameter as \(k\), \(k\)-max pooling selects the sub-matrix \({\bf{\bar C}} \in {\mathbb{R}^{  {(d_w/2)} \times {k} }}\) of the \(k\) highest values in each row of the matrix \({\bf{\hat C}}\). For dynamic \(k\)-max pooling, the pooling parameter \(k\) is dynamically selected in order to allow for a smooth extraction of higher-order and longer-range features~\cite{16_blunsom2014convolutional}. Given a fixed pooling parameter \(k_{top}\) for the topmost convolutional layer, the parameter \(k\) of \(k\)-max pooling in the \(l\)-th convolutional layer can be computed as follows:
\begin{equation}
{k_l} = \max ({k_{top}},\left\lceil {\frac{{L - l}}{L}s} \right\rceil ),
\label{eq:Ktop}
\end{equation}
where \(L\) is the total number of convolutional layers in the network.

\subsection{Unsupervised Dimensionality Reduction}
\label{sec:Unsupervised}
As described in Figure~\ref{fig:Architecture}, the dimensionality reduction function is defined as follows:
\begin{equation}
{\bf{Y}} = {f_{dr}}({\bf{X}}),
\end{equation}
where, \({\bf{Y}} \in {\mathbb{R}^{q \times n}}\) are the \(q\)-dimensional reduced latent space representations. Here, we take four popular dimensionality reduction methods as examples in our framework.

{\bf{Average Embedding (AE)}}: This method directly averages the word embeddings which are respectively weighted with TF and TF-IDF. Huang et al.~\cite{13_huang2012improving} used this strategy as the global context in their task, and Socher et al.~\cite{35_socher2013recursive} and Lai et al.~\cite{14_lai2015rcnn} used this method for text classification. The weighted average of all word vectors in one text can be computed as follows:

\begin{equation}
{\bf{Y}}({x_i}) = \frac{{\sum\nolimits_{i = 1}^k {{\bf{w}}({w_i}) \cdot {\bf{e}}({w_i})} }}{{\sum\nolimits_{i = 1}^k {{\bf{w}}({w_i})} }},
\end{equation}
where \({{\bf{w}}({w_i})}\) can be any weighting function that captures the importance of word \({{w_i}}\) in the text \({{\bf {x}}}\).

{\bf{Latent Semantic Analysis (LSA)}}: LSA~\cite{deerwester1990indexing} is the most popular global matrix factorization method, which applies a dimension reducing linear projection, Singular Value Decomposition (SVD), of the corresponding term/document matrix. Suppose the rank of \({\bf{X}}\) is \({ \hat{r}} \), LSA decompose \({\bf{X}}\) into the product of three other matrices:

\begin{equation}
{\bf{X}} = {\bf{U\Sigma }}{{\bf{V}}^T},
\end{equation}
where \({\bf{\Sigma }} = diag({\sigma _1},...,{\sigma _{\hat{r}}})\) and \({\sigma _1} \ge {\sigma _2} \ge ... \ge {\sigma _{\hat{r}}}\) are the singular values of \({\bf{X}}\), \({\bf{U}} \in {\mathbb{R}^{d \times {\hat{r}}}}\) is a set of left singular vectors and \({\bf{V}} \in {\mathbb{R}^{n \times {\hat{r}}}}\) is a set of right singular vectors. LSA uses the top \( q \) vectors in \( {\bf {U}}\) as the transformation matrix to embed the original text features into a \( q \)-dimensional subspace \( {\bf {Y}}\)~\cite{deerwester1990indexing}.

{\bf{Laplacian Eigenmaps (LE)}}: The top eigenvectors of graph Laplacian, defined on the similarity matrix of texts, are used in the method, which can discover the manifold structure of the text space~\cite{37_ng2002spectral}. In order to avoid storing the dense similarity matrix, many approximation techniques are proposed to reduce the memory usage and computational complexity for LE. There are two representative approximation methods, sparse similarity matrix and Nystr\({\ddot{\text{o}}}\)m approximation. Following previous studies~\cite{4_cai2005document,28_zhang2010self}, we select the former technique to construct the \(n \times n\) local similarity matrix \({\bf{A}}\) by using heat kernel as follows:

\begin{equation}
{A_{ij}}\!\! =\!\! \left\{ {\begin{array}{*{20}{c}}
{\!\!\!\!\exp (\! - \frac{{{{\left\| {{{\bf{x}}_i} - {{\bf{x}}_j}} \right\|}^2}}}{{2{\sigma ^2}}}\!),\!\!\!\!}&{i\!f~{\rm{ }}{{\bf{x}}_i}\!\! \in \!\! {{\bf{N}}_k}({\bf{x}}_j^{}){\rm{ }}~or{\rm{ }}~vice~{\rm{ versa}}}\\
{\!\!\!\!0,}&\!\!\!\!{otherwise}
\end{array}} \right.
\label{eq:SimilarityFunction}
\end{equation}
where, \(\sigma \) is a tuning parameter (default is 1) and \({{\bf{N}}_k}({\bf{x}})\) represents the set of \(k\)-nearest-neighbors of \({\bf{x}}\). By introducing a diagonal \(n \times n\) matrix \({\bf{D}}\) whose entries are given by \({D_{ii}} = \sum\nolimits_{j = 1}^n {{A_{ij}}} \), the graph Laplacian \({\bf{L}}\) can be computed by (\({\bf{D}}-{\bf{A}}\)). The optimal \(q \times n\) real-valued matrix \({\bf{ Y}}\) can be obtained by solving the following objective function:
\begin{equation}
\begin{array}{*{20}{c}}
{\mathop {\arg \min }\limits_{{\bf{ Y}}} }&{}&{tr({{{\bf{ Y}}}}{\bf{L Y^T}})}\\
{{\rm{subject~to}}}&{}&{{{{\bf{ Y}}}}{\bf{D Y^T}} = {\bf{I}}}\\
{}&{}&{{{{\bf{ Y}}}}{\bf{D1}} = {\bf{0}}}
\end{array}
\end{equation}
where \(tr( \cdot )\) is the trace function, \({{\bf{ Y}}}{\bf{D Y^T}} = {\bf{I}}\) requires the different dimensions to be uncorrelated, and \({{\bf{ Y}}}{\bf{D1}} = {\bf{0}}\) requires each dimension to achieve equal probability as positive or negative).

{\bf{Locality Preserving Indexing (LPI)}}: This method extends LE to deal with unseen texts by approximating the linear function \({\bf{Y}} = {\bf{W}}_{LPI}^T{\bf{X}}\)~\cite{28_zhang2010self}, and the subspace vectors are obtained by finding the optimal linear approximations to the eigenfunctions of the Laplace Beltrami operator on the Riemannian manifold~\cite{niyogi2004locality}. Similar as LE, we first construct the local similarity matrix \({\bf{A}}\), then the graph Laplacian \({\bf{L}}\) can be computed by (\({\bf{D}}-{\bf{A}}\)), where \({D_{ii}} \) measures the local density around \({{\bf{x}}_{i}} \) and is equal to \( \sum\nolimits_{j = 1}^n {{A_{ij}}} \). Compute the eigenvectors \({\bf{a}}\) and eigenvalues \({\bf{\lambda }}\) of the following generalized eigen-problem:

\begin{equation}
{\bf{XL}}{{\bf{X}}^T}{\bf{a}} = {\bf{\lambda XD}}{{\bf{X}}^T}{\bf{a}}.
\end{equation}

The mapping function \({\bf{W}}_{LPI}^{} = [{{\bf{a}}_1},...,{{\bf{a}}_q}]\) can be obtained and applied to the unseen data~\cite{4_cai2005document}.

All of the above methods claim a better performance in capturing semantic similarity between texts in the reduced latent space representation \({\bf{Y}}\) than in the original representation \({\bf{X}}\), while the performance of short text clustering can be further enhanced with the help of our framework, self-taught CNN.

\subsection{Learning}
The last layer of CNN is an output layer as follows:
\begin{equation}
{{\bf{O}}} = {{\bf{W}}_O}{\bf{h}},
\label{eq:Output}
\end{equation}
where, \({\bf{h}}\) is the deep feature representation, \({\bf{O}} \in {\mathbb{R}^q}\) is the output vector and \({{\bf{W}}_O} \in {\mathbb{R}^{q \times r}}\) is weight matrix.

In order to incorporate the latent semantic features \({\bf{Y}}\), we first binary the real-valued vectors \({\bf{Y}}\) to the binary codes \({\bf{B}}\) by setting the threshold to be the media vector \(median({\bf{Y}})\). Then, the output vector \({\bf{O}}\) is used to fit the binary codes \({\bf{B}}\) via \(q\) logistic operations as follows:
\begin{equation}
p_i^{} = \frac{{\exp ({\bf{O}}_i^{})}}{{1 + \exp ({\bf{O}}_i^{})}}.
\label{eq:logistic}
\end{equation}

All parameters to be trained are defined as \(\theta \).
\begin{equation}
\theta  = \{ {\bf{E}},{\bf{\tilde W}},{{\bf{W}}_O}\}.
\label{eq:theta}
\end{equation}

Given the training text collection \({\bf{X}}\), and the pre-trained binary codes \({\bf{B}}\), the log likelihood of the parameters can be written down as follows:
\begin{equation}
{J({\bf{\theta }})} = \sum\limits_{i = 1}^n {\log p({{\bf{b}}_i}|{{\bf{x}}_i},{\bf{\theta }})}.
\label{eq:likelihood}
\end{equation}

Following the previous work~\cite{16_blunsom2014convolutional}, we train the network with mini-batches by back-propagation and perform the gradient-based optimization using the Adagrad update rule~\cite{36_duchi2011adaptive}. For regularization, we employ dropout with 50\% rate to the penultimate layer~\cite{16_blunsom2014convolutional,22_kim2014convolutional}.

\subsection{K-means for Clustering}
With the given short texts, we first utilize the trained deep neural network to obtain the semantic representations \({\bf{h}}\), and then employ traditional K-means algorithm to perform clustering.

\section{Experiments}
\label{sec:Experiments}

\begin{table}[t] 
\begin{center}
\begin{tabular}{|c|c|c|c|c|}\hline
Dataset&C &Num. &Len. &$|V|$\\\hline \hline
SearchSnippets   &	8    &	12,340     &	17.88/38  &	30,642       \\
StackOverflow     &	20    &	20,000     &	8.31/34  &	22,956       \\
Biomedical     &	20    &	20,000     &	 12.88/53 &	18,888       \\\hline
\end{tabular}
\end{center}
\caption{\label{tb:Datasets} Statistics for the text datasets. C: the number of classes; Num: the dataset size; Len.: the mean/max length of texts and $|V|$: the vocabulary size.}
\end{table}

\begin{table}[t] 
\begin{center}
\begin{tabular}{|c|c|c|c|}\hline
\multicolumn{4}{|l|}{SearchSnippets: 8 different domains}\\\hline
 business & computers & health & education  \\\hline
culture  & engineering& sports& politics  \\\hline\hline

\multicolumn{4}{|l|}{StackOverflow: 20 semantic tags}\\\hline
 svn & oracle & bash & apache  \\\hline
excel & matlab& cocoa& visual-studio  \\\hline
 osx & wordpress& spring &hibernate \\\hline
 scala & sharepoint & ajax & drupal \\\hline
qt  & haskell  & linq & magento \\\hline\hline

\multicolumn{4}{|l|}{Biomedical: 20 MeSH major topics}\\\hline
 aging & chemistry& cats  & erythrocytes  \\\hline
glucose & potassium& lung & lymphocytes  \\\hline
spleen & mutation&skin  &norepinephrine \\\hline
insulin & prognosis & risk & myocardium \\\hline
sodium  &  mathematics & swine & temperature \\\hline
\end{tabular}
\end{center}
\caption{\label{tb:SemanticTopics} Description of semantic topics (that is, tags/labels) from the three text datasets used in our experiments.}
\end{table}

\subsection{Datasets}
We test our proposed approach on three public short text datasets. The summary statistics and semantic topics of these datasets are described in Table~\ref{tb:Datasets} and Table~\ref{tb:SemanticTopics}.

{\bf{SearchSnippets}}\footnote{\url{http://jwebpro.sourceforge.net/data-web-snippets.tar.gz}.}. This dataset was selected from the results of web search transaction using predefined phrases of 8 different domains by Phan et al.~\cite{20_phan2008learning}.

{\bf{StackOverflow}}. We use the challenge data published in Kaggle.com\footnote{\url{https://www.kaggle.com/c/predict-closed-questions-on-stack-overflow/download/train.zip}.}. The raw dataset consists 3,370,528 samples through July 31st, 2012 to August 14, 2012. In our experiments, we randomly select 20,000 question titles from 20 different tags as in Table~\ref{tb:SemanticTopics}.

{\bf{Biomedical}}. We use the challenge data published in BioASQ's official website\footnote{\url{http://participants-area.bioasq.org/}.}. In our experiments, we randomly select 20, 000 paper titles from 20 different MeSH\footnote{\url{http://en.wikipedia.org/wiki/Medical_Subject_Headings}.} major topics as in Table~\ref{tb:SemanticTopics}. As described in Table~\ref{tb:Datasets}, the max length of selected paper titles is 53\footnote{\url{http://www.ncbi.nlm.nih.gov/pubmed/207752}.}.

For these datasets, we randomly select 10\% of data as the development set. Since SearchSnippets has been pre-processed by Phan et al.~\cite{20_phan2008learning}, we do not further process this dataset. In StackOverflow, texts contain lots of computer terminology, and symbols and capital letters are meaningful, thus we do not do any pre-processed procedures. For Biomedical, we remove the symbols and convert letters into lower case.

\begin{table}[t] 
\begin{center}
\begin{tabular}{|c|c|c|}\hline
      Dataset           &	$|V|$  &	$|T|$       \\\hline \hline
SearchSnippets&23,826 (77\%) &	211,575 (95\%)\\
StackOverflow&	19,639 (85\%)  &	162,998 (97\%) \\
Biomedical&18,381 (97\%) & 257,184 (99\%)\\\hline
\end{tabular}
\end{center}
\caption{\label{tb:Embeddings} Coverage of word embeddings on three datasets. $|V|$ is the vocabulary size and $|T|$ is the number of tokens.}
\end{table}

\subsection{Pre-trained Word Vectors}
We use the publicly available word2vec\footnote{\url{https://code.google.com/p/word2vec/}.} tool to train word embeddings, and the most parameters are set as same as Mikolov et al.~\cite{21_mikolov2013distributed} to train word vectors on Google News setting\footnote{\url{https://groups.google.com/d/msg/word2vec-toolkit/lxbl_MB29Ic/NDLGId3KPNEJ}.}, except of vector dimensionality using 48 and minimize count using 5. For SearchSnippets, we train word vectors on Wikipedia dumps\footnote{\url{http://dumps.wikimedia.org/enwiki/latest/enwiki-latest-pages-articles.xml.bz2}.}. For StackOverflow, we train word vectors on the whole corpus of the StackOverflow dataset described above which includes the question titles and post contents. For Biomedical, we train word vectors on all titles and abstracts of 2014 training articles. The coverage of these learned vectors on three datasets are listed in Table~\ref{tb:Embeddings}, and the words not present in the set of pre-trained words are initialized randomly.

\subsection{Comparisons}
\label{sec:Comparisons}
In our experiment, some widely used text clustering methods are compared with our approach. Besides K-means, Skip-thought Vectors, Recursive Neural Network and Paragraph Vector based clustering methods, four baseline clustering methods are directly based on the popular unsupervised dimensionality reduction methods as described in Section~\ref{sec:Unsupervised}. We further compare our approach with some other non-biased neural networks, such as bidirectional RNN. More details are listed as follows:

{\bf{K-means}} K-means~\cite{2_wagstaff2001constrained} on original keyword features which are respectively weighted with term frequency (TF) and term frequency-inverse document frequency (TF-IDF).

{\bf{Skip-thought Vectors (SkipVec)}} This baseline~\cite{kiros2015skip} gives an off-the-shelf encoder to produce highly generic sentence representations. The encoder\footnote{\url{https://github.com/ryankiros/skip-thoughts}.} is trained using a large collection of novels and provides three encoder modes, that are unidirectional encoder (SkipVec (Uni)) with 2,400 dimensions, bidirectional encoder (SkipVec (Bi)) with 2,400 dimensions and combined encoder (SkipVec (Combine)) with SkipVec (Uni) and SkipVec (Bi) of 2,400 dimensions each. K-means is employed on the these vector representations respectively.

 {\bf{Recursive Neural Network (RecNN)}} In~\cite{24_socher2011semi}, the tree structure is firstly greedy approximated via unsupervised recursive autoencoder. Then, semi-supervised recursive autoencoders are used to capture the semantics of texts based on the predicted structure. In order to make this recursive-based method completely unsupervised, we remove the cross-entropy error in the second phrase to learn vector representation and subsequently employ K-means on the learned vectors of the top tree node and the average of all vectors in the tree.

 {\bf{Paragraph Vector (Para2vec)}} K-means on the fixed size feature vectors generated by Paragraph Vector (Para2vec)~\cite{le2014distributed} which is an unsupervised method to learn distributed representation of words and paragraphs. In our experiments, we use the open source software\footnote{\url{https://github.com/mesnilgr/iclr15}.} released by Mesnil et al.~\cite{mesnil2014ensemble}.

 {\bf{Average Embedding (AE)}} K-means on the weighted average vectors of the word embeddings which are respectively weighted with TF and TF-IDF. The dimension of average vectors is equal to and decided by the dimension of word vectors used in our experiments.

 {\bf{Latent Semantic Analysis (LSA)}} K-means on the reduced subspace vectors generated by Singular Value Decomposition (SVD) method. The dimension of subspace is default set to the number of clusters, we also iterate the dimensions ranging from 10:10:200 to get the best performance, that is 10 on SearchSnippets, 20 on StackOverflow and 20 on Biomedical in our experiments.

 {\bf{Laplacian Eigenmaps (LE)}} This baseline, using Laplacian Eigenmaps and subsequently employing K-means algorithm, is well known as spectral clustering~\cite{12_belkin2001laplacian}. The dimension of subspace is default set to the number of clusters~\cite{37_ng2002spectral,4_cai2005document}, we also iterate the dimensions ranging from 10:10:200 to get the best performance, that is 20 on SearchSnippets, 70 on StackOverflow and 30 on Biomedical in our experiments.

 {\bf{Locality Preserving Indexing (LPI)}} This baseline, projecting the texts into a lower dimensional semantic space, can discover both the geometric and discriminating structures of the original feature space~\cite{4_cai2005document}. The dimension of subspace is default set to the number of clusters~\cite{4_cai2005document}, we also iterate the dimensions ranging from 10:10:200 to get the best performance, that is 20 on SearchSnippets, 80 on StackOverflow and 30 on Biomedical in our experiments.

{\bf{bidirectional RNN (bi-RNN)}} We replace the CNN model in our framework as in Figure~\ref{fig:Architecture} with some bi-RNN models. Particularly, LSTM and GRU units are used in the experiments. In order to generate the fixed-length document representation from the variable-length vector sequences, for both bi-LSTM and bi-GRU based clustering methods, we further utilize three pooling methods:  last pooling (using the last hidden state), mean pooling and element-wise max pooling. These pooling methods are respectively used in the previous works~\cite{palangi2015deep,cho2014learning}, \cite{tang2015document} and \cite{14_lai2015rcnn}. For regularization, the training gradients of all parameters with an \(l\)2 norm larger than 40 are clipped to 40, as the previous work~\cite{sukhbaatar2015end}.

\subsection{Evaluation Metrics}
The clustering performance is evaluated by comparing the clustering results of texts with the tags/labels provided by the text corpus. Two metrics, the accuracy (ACC) and the normalized mutual information metric (NMI), are used to measure the clustering performance~\cite{4_cai2005document,1_huang2014deep}. Given a text \({{\bf{x}}_i}\), let \({c_i}\) and \({t_i}\) be the obtained cluster label and the label provided by the corpus, respectively.
Accuracy is defined as:
\begin{equation}
ACC = \frac{{\sum\nolimits_{i = 1}^n {\delta ({t_i},map({c_i}))} }}{n},
\label{eq:ACC}
\end{equation}
where, \(n\) is the total number of texts, \(\delta (x,y)\) is the indicator function that equals one if \(x = y\) and equals zero otherwise, and \(map({c_i})\) is the permutation mapping function that maps each cluster label \({c_i}\) to the equivalent label from the text data by Hungarian algorithm~\cite{7_papadimitriou1998combinatorial}.

Normalized mutual information~\cite{8_chen2011parallel} between tag/label set \({\bf{T}}\) and cluster set \({\mathbb{C}}\) is a popular metric used for evaluating clustering tasks. It is defined as follows:
\begin{equation}
NMI({\bf{T}},{\mathbb{C}}) = \frac{{MI({\bf{T}},{\mathbb{C}})}}{{\sqrt {H({\bf{T}})H({\mathbb{C}})} }},
\label{eq:NMI}
\end{equation}
where, \(MI({\bf{T}},{\mathbb{C}})\) is the mutual information between \({\bf{T}}\) and \({\mathbb{C}}\), \(H( \cdot )\) is entropy and the denominator \(\sqrt {H({\bf{T}})H({\mathbb{C}})} \) is used for normalizing the mutual information to be in the range of [0, 1].

\begin{table*}[t] 
\begin{center}
\begin{tabular}{|l|c|c|c|}\hline
&SearchSnippets &StackOverflow &Biomedical\\\hline
Method     &~~~~ACC (\%)~~~~&~~~~ACC (\%)~~~~&~~~~ACC (\%)~~~~\\\hline \hline
K-means (TF)& 24.75$\pm$2.22        & 13.51$\pm$2.18  & 15.18$\pm$1.78    \\
K-means (TF-IDF)& 33.77$\pm$3.92    & 20.31$\pm$3.95  & 27.99$\pm$2.83 \\
SkipVec (Uni) & 28.23$\pm$1.08           & 08.79$\pm$0.19  & 16.44$\pm$0.50 \\
SkipVec (Bi)  & 29.24$\pm$1.57           & 09.59$\pm$0.15  & 16.11$\pm$0.60 \\
SkipVec (Combine)  & 33.58$\pm$1.95      & 09.34$\pm$0.24  & 16.27$\pm$0.33 \\
RecNN (Top) & 21.21$\pm$1.62     	&13.13$\pm$0.80	  &13.73$\pm$0.67\\
RecNN (Ave.) & 65.59$\pm$5.35		&40.79$\pm$1.38	  &37.05$\pm$1.27\\
RecNN (Top+Ave.) & 65.53$\pm$5.64	&40.45$\pm$1.60	  &36.68$\pm$1.29\\
Para2vec & 69.07$\pm$2.53	&32.55$\pm$0.89	  &41.26$\pm$1.22\\
\hline
STC$^2$-AE& 68.34$\pm$2.51& 40.05$\pm$1.77	  &37.44$\pm$1.19\\
STC$^2$-LSA& 73.09$\pm$1.45	& 35.81$\pm$1.80	   &38.47$\pm$1.55\\
STC$^2$-LE& {\bf{77.09$\pm$3.99}} & {\bf{51.13$\pm$2.80}}	     &{\bf{43.62$\pm$1.00}}\\
STC$^2$-LPI& {\bf{77.01$\pm$4.13}}	& {\bf{51.14$\pm$2.92}}	&43.00$\pm$1.25\\\hline

\end{tabular}
\end{center}
\caption{\label{tb:Comparison} Comparison of ACC of our proposed methods and three clustering methods on three datasets. For RecNN (Top), K-means is conducted on the learned vectors of the top tree node. For RecNN (Ave.), K-means is conducted on the average of all vectors in the tree. More details about the baseline setting are described in Section~\ref{sec:Comparisons}}
\end{table*}

\begin{table*}[t] 
\begin{center}
\begin{tabular}{|l|c|c|c|}\hline
&SearchSnippets &StackOverflow &Biomedical\\\hline
Method      &~~~~NMI (\%)~~~~&~~~~NMI (\%)~~~~&~~~~NMI (\%)~~~~\\\hline \hline
K-means (TF)   &	09.03$\pm$2.30    & 07.81$\pm$2.56  &09.36$\pm$2.04    \\
K-means (TF-IDF)   &	21.40$\pm$4.35    & 15.64$\pm$4.68   &25.43$\pm$3.23 \\
SkipVec (Uni) & 10.98$\pm$0.93           & 02.24$\pm$0.13  & 10.52$\pm$0.41 \\
SkipVec (Bi)  & 09.27$\pm$0.29           & 02.89$\pm$0.20  & 10.15$\pm$0.59 \\
SkipVec (Combine)  & 13.85$\pm$0.78      & 02.72$\pm$0.34  & 10.72$\pm$0.46 \\
RecNN (Top) 	&04.04$\pm$0.74		&09.90$\pm$0.96   &08.87$\pm$0.53\\
RecNN (Ave.) 	&50.55$\pm$1.71		&40.58$\pm$0.91   &33.85$\pm$0.50\\
RecNN (Top+Ave.) &50.44$\pm$1.84	&40.21$\pm$1.18   &33.75$\pm$0.50\\
Para2vec & 50.51$\pm$0.86	&27.86$\pm$0.56	  &34.83$\pm$0.43\\
\hline
STC$^2$-AE&	54.01$\pm$1.55 & 38.22$\pm$1.31 &33.58$\pm$0.48\\
STC$^2$-LSA&	54.53$\pm$1.47	& 34.38$\pm$1.12   &33.90$\pm$0.67\\
STC$^2$-LE&	{\bf{63.16$\pm$1.56}}	& {\bf{49.03$\pm$1.46}}  &38.05$\pm$0.48\\
STC$^2$-LPI&	62.94$\pm$1.65	& {\bf{49.08$\pm$1.49}} &{\bf{38.18$\pm$0.47}}\\\hline

\end{tabular}
\end{center}
\caption{\label{tb:ComparisonNMI} Comparison of NMI of our proposed methods and three clustering methods on three datasets. For RecNN (Top), K-means is conducted on the learned vectors of the top tree node. For RecNN (Ave.), K-means is conducted on the average of all vectors in the tree. More details about the baseline setting are described in Section~\ref{sec:Comparisons}}
\end{table*}

\subsection{Hyperparameter Settings}
The most of parameters are set uniformly for these datasets. Following previous study~\cite{4_cai2005document}, the number of nearest neighbors in Eqn.~(\ref{eq:SimilarityFunction}) is fixed to 15 when constructing the graph structures for LE and LPI. For CNN model, the networks has two convolutional layers. The widths of the convolutional filters are both 3. The value of \(k\) for the top \(k\)-max pooling in Eqn.~(\ref{eq:Ktop}) is 5. The number of feature maps at the first convolutional layer is 12, and 8 feature maps at the second convolutional layer. Both those two convolutional layers are followed by a folding layer. We further set the dimension of word embeddings \({d_w}\) as 48. Finally, the dimension of the deep feature representation \(r\) is fixed to 480. Moreover, we set the learning rate \(\lambda \) as 0.01 and the mini-batch training size as 200. The output size \(q\) in Eqn.~(\ref{eq:Output}) is set same as the best dimensions of subspace in the baseline method, as described in Section~\ref{sec:Comparisons}.

For initial centroids have significant impact on clustering results when utilizing the K-means algorithms, we repeat K-means for multiple times with random initial centroids (specifically, 100 times for statistical significance) as Huang~\cite{1_huang2014deep}. The all subspace vectors are normalized to 1 before applying K-means and the final results reported are the average of 5 trials with all clustering methods on three text datasets.

\begin{table*}[t] 
\begin{center}
\begin{tabular}{|l|c|c|c|}\hline
&SearchSnippets &StackOverflow &Biomedical\\\hline
Method     &~~~~ACC (\%)~~~~&~~~~ACC (\%)~~~~&~~~~ACC (\%)~~~~\\\hline \hline
bi-LSTM (last)& 64.50$\pm$3.18        & 46.83$\pm$1.79  & 36.50$\pm$1.08    \\
bi-LSTM (mean)& 65.85$\pm$4.18    & 44.93$\pm$1.83  & 35.60$\pm$1.21 \\
bi-LSTM (max) & 61.70$\pm$5.10     	&38.74$\pm$1.62	  &32.83$\pm$0.73\\
bi-GRU (last)& 70.18$\pm$2.62        & 43.36$\pm$1.46  & 35.19$\pm$0.78    \\
bi-GRU (mean)& 70.29$\pm$2.61    & 44.53$\pm$1.81  & 36.75$\pm$1.21 \\
bi-GRU (max) & 65.69$\pm$1.02     &	{\bf{54.40$\pm$2.07}}	  &37.23$\pm$1.19\\
\hline
LPI (best)& 47.11$\pm$2.91	& 38.04$\pm$1.72  &37.15$\pm$1.16\\
STC$^2$-LPI& {\bf{77.01$\pm$4.13}}	& 51.14$\pm$2.92	&{\bf{43.00$\pm$1.25}}\\\hline

\end{tabular}
\end{center}
\caption{\label{tb:Comparison_BIRNN} Comparison of ACC of our proposed methods and some other non-biased models on three datasets. For LPI, we project the text under the best dimension as described in Section~\ref{sec:Comparisons}. For both bi-LSTM and bi-GRU based clustering methods, the binary codes generated from LPI are used to guide the learning of bi-LSTM/bi-GRU models.}
\end{table*}

\begin{table*}[t] 
\begin{center}
\begin{tabular}{|l|c|c|c|}\hline
&SearchSnippets &StackOverflow &Biomedical\\\hline
Method      &~~~~NMI (\%)~~~~&~~~~NMI (\%)~~~~&~~~~NMI (\%)~~~~\\\hline \hline
bi-LSTM (last)& 50.32$\pm$1.15        & 41.89$\pm$0.90  & 34.51$\pm$0.34    \\
bi-LSTM (mean)& 52.11$\pm$1.69    & 40.93$\pm$0.91  & 34.03$\pm$0.28 \\
bi-LSTM (max) & 46.81$\pm$2.38     	&36.73$\pm$0.56	  &31.90$\pm$0.23\\
bi-GRU (last)& 56.00$\pm$0.75        & 38.73$\pm$0.78  & 32.91$\pm$0.40    \\
bi-GRU (mean)& 55.76$\pm$0.85    & 39.84$\pm$0.94  & 34.27$\pm$0.27 \\
bi-GRU (max) & 51.11$\pm$1.06     &	{\bf{51.10$\pm$1.31}}	  &32.74$\pm$0.34\\
\hline
LPI (best)&  38.48$\pm$2.39	& 27.21$\pm$0.88        &29.73$\pm$0.30\\
STC$^2$-LPI&	{\bf{62.94$\pm$1.65}}	& 49.08$\pm$1.49 &{\bf{38.18$\pm$0.47}}\\\hline

\end{tabular}
\end{center}
\caption{\label{tb:ComparisonNMI_BIRNN} Comparison of NMI of our proposed methods and some other non-biased models on three datasets. For LPI, we project the text under the best dimension as described in Section~\ref{sec:Comparisons}. For both bi-LSTM and bi-GRU based clustering methods, the binary codes generated from LPI are used to guide the learning of bi-LSTM/bi-GRU models.}
\end{table*}

\subsection{Results and Analysis}
In Table~\ref{tb:Comparison} and Table~\ref{tb:ComparisonNMI}, we report the ACC and NMI performance of our proposed approaches and four baseline methods, K-means, SkipVec, RecNN and Para2vec based clustering methods. Intuitively, we get a general observation that (1) BoW based approaches, including K-means (TF) and K-means (TF-IDF), and SkipVec based approaches perform not well; (2) RecNN based approaches, both RecNN (Ave.) and RecNN (Top+Ave.), do better; (3) Para2vec makes a comparable performance with the most baselines; and (4) the evaluation clearly demonstrate the superiority of our proposed methods STC$^2$. It is an expected results.
For SkipVec based approaches, the off-the-shelf encoders are trained on the BookCorpus datasets~\cite{zhu2015aligning}, and then applied to our datasets to extract the sentence representations. The SkipVec encoders can produce generic sentence representations but may not perform well for specific datasets, in our experiments, StackOverflow and Biomedical datasets consist of many computer terms and medical terms, such as ``ASP.NET'', ``XML'', ``C\#'', ``serum'' and ``glycolytic''.
When we take a more careful look, we find that RecNN (Top) does poorly, even worse than K-means (TF-IDF). The reason maybe that although recursive neural models introduce tree structure to capture compositional semantics, the vector of the top node mainly captures a biased semantic while the average of all vectors in the tree nodes, such as RecNN (Ave.), can be better to represent sentence level semantic. And we also get another observation that, although our proposed STC$^2$-LE and STC$^2$-LPI outperform both BoW based and RecNN based approaches across all three datasets, STC$^2$-AE and STC$^2$-LSA do just exhibit some similar performances as RecNN (Ave.) and RecNN (Top+Ave.) do in the datasets of StackOverflow and Biomedical.

We further replace the CNN model in our framework as in Figure~\ref{fig:Architecture} with some other non-biased models, such as bi-LSTM and bi-GRU, and report the results in Table~\ref{tb:Comparison_BIRNN} and Table~\ref{tb:ComparisonNMI_BIRNN}. As an instance, the binary codes generated from LPI are used to guide the learning of bi-LSTM/bi-GRU models. From the results, we can see that bi-GRU and bi-LSTM based clustering methods do equally well, no clear winner, and both achieve great enhancements compared with LPI (best). Compared with these bi-LSTM/bi-GRU based models, the evaluation results still demonstrate the superiority of our approach methods, CNN based clustering model, in the most cases. As the results reported by Visin et al.~\cite{visin2015renet}, despite bi-directional or multi-directional RNN models perform a good non-biased feature extraction, they yet do not outperform state-of-the-art CNN on some tasks.

\begin{figure}[t]
\centering
\vspace{-0.3cm}
\subfigure{
\includegraphics[width=7.7cm]{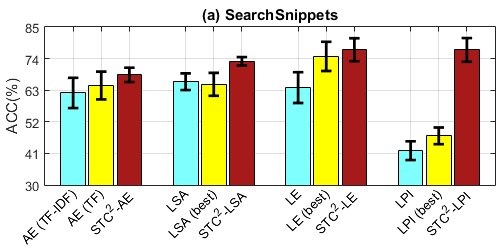}}
\vspace{-0.3cm}
\subfigure{
\includegraphics[width=7.7cm]{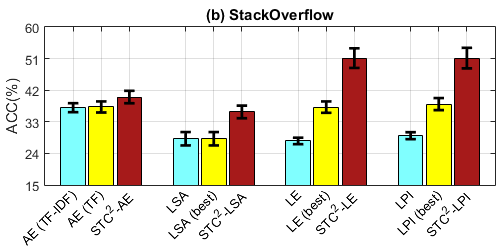}}
\vspace{-0.3cm}
\subfigure{
\includegraphics[width=7.7cm]{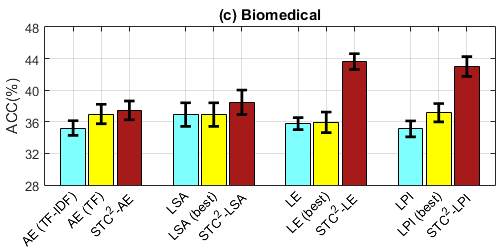}}
\caption{ACC results on three short text datasets using our proposed STC$^2$ based on AE, LSA, LE and LPI.}
\vspace{-0.3cm}
\label{fig:ACCBar}
\end{figure}

\begin{figure}[t]
\centering
\vspace{-0.3cm}
\subfigure{
\includegraphics[width=7.7cm]{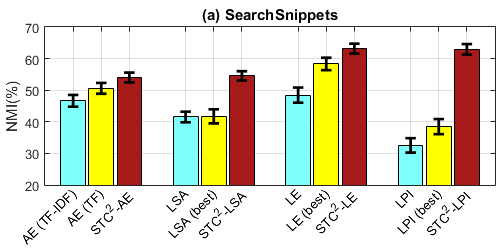}}
\vspace{-0.3cm}
\subfigure{
\includegraphics[width=7.7cm]{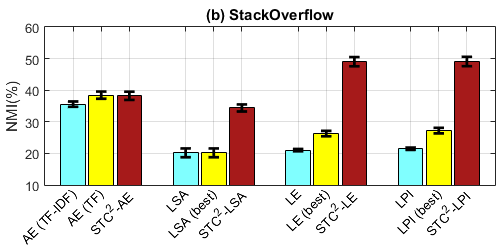}}
\vspace{-0.3cm}
\subfigure{
\includegraphics[width=7.7cm]{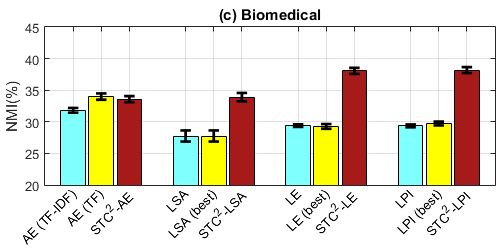}}
\caption{NMI results on three short text datasets using our proposed STC$^2$ based on AE, LSA, LE and LPI.}
\vspace{-0.3cm}
\label{fig:NMIBar}
\end{figure}

In order to make clear what factors make our proposed method work, we report the bar chart results of ACC and MNI of our proposed methods and the corresponding baseline methods in Figure~\ref{fig:ACCBar} and Figure~\ref{fig:NMIBar}. It is clear that, although AE and LSA does well or even better than LE and LPI, especially in dataset of both StackOverflow and Biomedical, STC$^2$-LE and STC$^2$-LPI achieve a much larger performance enhancements than STC$^2$-AE and STC$^2$-LSA do. The possible reason is that the information the pseudo supervision used to guide the learning of CNN model that make difference. Especially, for AE case, the input features fed into CNN model and the pseudo supervision employed to guide the learning of CNN model are all come from word embeddings. There are no different semantic features to be used into our proposed method, thus the performance enhancements are limited in STC$^2$-AE. For LSA case, as we known, LSA is to make matrix factorization to find the best subspace approximation of the original feature space to minimize the global reconstruction error. And as~\cite{26_pennington2014glove,li2015mfp} recently point out that word embeddings trained with word2vec or some variances, is essentially to do an operation of matrix factorization. Therefore, the information between input and the pseudo supervision in CNN is not departed very largely from each other, and the performance enhancements of STC$^2$-AE is also not quite satisfactory. For LE and LPI case, as we known that LE extracts the manifold structure of the original feature space, and LPI extracts both geometric and discriminating structure of the original feature space~\cite{4_cai2005document}. We guess that our approach STC$^2$-LE and STC$^2$-LPI achieve enhancements compared with both LE and LPI by a large margin, because both of LE and LPI get useful semantic features, and these features are also different from word embeddings used as input of CNN. From this view, we say that our proposed STC has potential to behave more effective when the pseudo supervision is able to get semantic meaningful features, which is different enough from the input of CNN.

\begin{figure}[h!]
\centering
\vspace{-0.1cm}
\vspace{-0.1cm}
\subfigure{
\includegraphics[width=3.81cm]{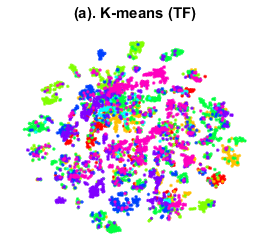}}
\vspace{-0.1cm}
\subfigure{
\includegraphics[width=3.81cm]{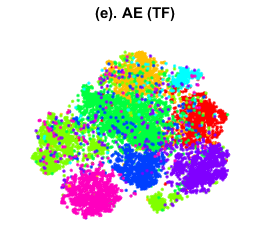}}
\vspace{-0.1cm}
\subfigure{
\includegraphics[width=3.81cm]{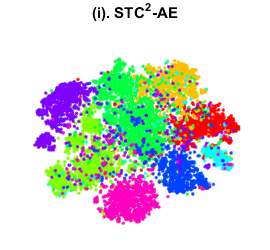}}
\vspace{-0.1cm}
\subfigure{
\includegraphics[width=3.81cm]{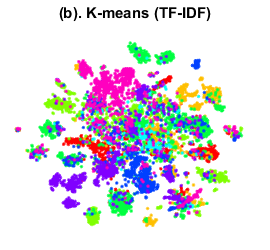}}
\vspace{-0.1cm}
\subfigure{
\includegraphics[width=3.81cm]{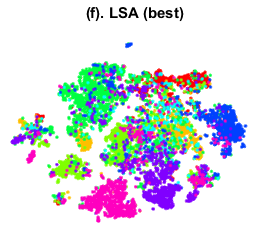}}
\vspace{-0.1cm}
\subfigure{
\includegraphics[width=3.81cm]{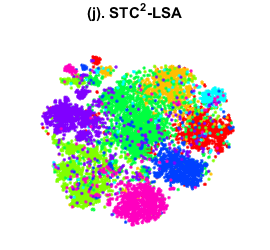}}
\vspace{-0.1cm}
\subfigure{
\includegraphics[width=3.81cm]{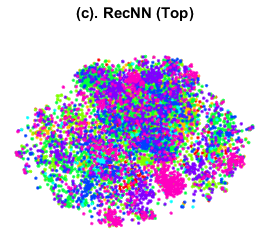}}
\vspace{-0.1cm}
\subfigure{
\includegraphics[width=3.81cm]{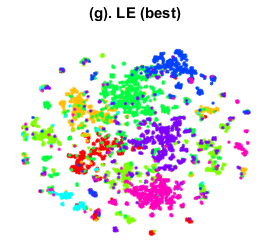}}
\vspace{-0.1cm}
\subfigure{
\includegraphics[width=3.81cm]{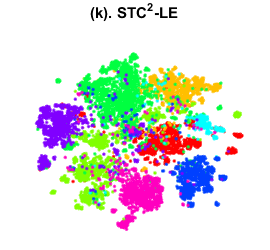}}
\vspace{-0.1cm}
\subfigure{
\includegraphics[width=3.81cm]{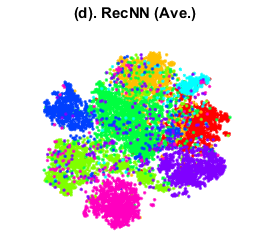}}
\vspace{-0.1cm}
\subfigure{
\includegraphics[width=3.81cm]{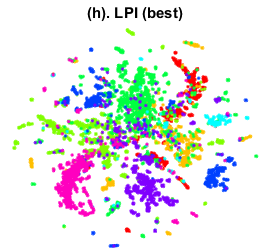}}
\vspace{-0.1cm}
\subfigure{
\includegraphics[width=3.81cm]{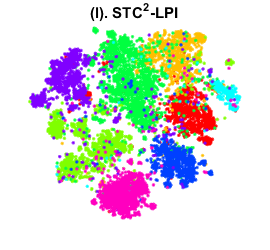}}
\vspace{-0.1cm}
\caption{A 2-dimensional embedding of original keyword features weighted with (a) TF and (b) TF-IDF, (c) vectors of the top tree node in RecNN, (d) average vectors of all tree node in RecNN, (e) average embeddings weighted with TF, subspace features based on (f) LSA, (g) LE and (h) LPI, deep learned features from (i) STC$^2$-AE, (j) STC$^2$-LSA, (k) STC$^2$-LE and (l) STC$^2$-LPI. All above features are respectively used in K-means (TF), K-means (TF-IDF), RecNN (Top), RecNN (Ave.), AE (TF), LSA(best), LE (best), LPI (best), and our proposed STC$^2$-AE, STC$^2$-LSA, STC$^2$-LE and STC$^2$-LPI on SearchSnippets. (Best viewed in color)}
\vspace{-0.2cm}
\label{fig:all2dSearchSnippets}
\end{figure}

Furthermore, from the results of K-means and AE in Table~\ref{tb:Comparison}-\ref{tb:ComparisonNMI} and Figure~\ref{fig:ACCBar}-\ref{fig:NMIBar}, we note that TF-IDF weighting gives a more remarkable improvement for K-means, while TF weighting works better than TF-IDF weighting for Average Embedding. Maybe the reason is that pre-trained word embeddings encode some useful information from external corpus and are able to get even better results without TF-IDF weighting. Meanwhile, we find that LE get quite unusual good performance than LPI, LSA and AE in SearchSnippets dataset, which is not found in the other two datasets. To get clear about this, and also to make a much better demonstration about our proposed approaches and other baselines, we further report 2-dimensional text embeddings on SearchSnippets in Figure~\ref{fig:all2dSearchSnippets}, using t-SNE\footnote{\url{http://lvdmaaten.github.io/tsne/}.}~\cite{39_van2008visualizing} to get distributed stochastic neighbor embedding of the feature representations used in the clustering methods.
We can see that the results of from AE and LSA seem to be fairly good or even better than the ones from LE and LPI, which is not the same as the results from ACC and NMI in Figure~\ref{fig:ACCBar}-\ref{fig:NMIBar}. Meanwhile, RecNN (Ave.) performs better than BoW (both TF and TF-IDF) while RecNN (Top) does not, which is the same as the results from ACC and NMI in Table~\ref{tb:Comparison} and Table~\ref{tb:ComparisonNMI}. Then we guess that both ''the same as'' and ''not the same as'' above, is just a good example to illustrate that visualization tool, such as t-SNE, get some useful information for measuring results, which is different from the ones of ACC and NMI. Moreover, from this complementary view of t-SNE, we can see that our STC$^2$-AE, STC$^2$-LSA, STC$^2$-LE, and STC$^2$-LPI show more clear-cut margins among different semantic topics (that is, tags/labels), compared with AE, LSA, LE and LPI, respectively, as well as compared with both baselines, BoW and RecNN based ones.

From all these results, with three measures of ACC, NMI and t-SNE under three datasets, we can get a solid conclusion that our proposed approaches is an effective approaches to get useful semantic features for short text clustering.

\section{Conclusions}
\label{sec:DiscussionANDConclusions}

With the emergence of social media, short text clustering has become an increasing important task. This paper explores a new perspective to cluster short texts based on deep feature representation learned from the proposed self-taught convolutional neural networks. Our framework can be successfully accomplished without using any external tags/labels and complicated NLP pre-processing, and and our approach is a flexible framework, in which the traditional dimension reduction approaches could be used to get performance enhancement. Our extensive experimental study on three short text datasets shows that our approach can achieve a significantly better performance. In the future, how to select and incorporate more effective semantic features into the proposed framework would call for more research.


\section*{Acknowledgments}
We would like to thank reviewers for their comments, and acknowledge Kaggle and BioASQ for making the datasets available. This work is supported by the National Natural Science Foundation of China (No. 61602479, No. 61303172, No. 61403385) and the Strategic Priority Research Program of the Chinese Academy of Sciences (Grant No. XDB02070005).

\section*{References}

\bibliography{mybibfile}

\end{document}